\documentclass[useAMS,usenatbib]{mn2e}

\usepackage{graphicx}

\title[Searching for IR excesses in Sun-like stars observed by {\em WISE}]{Searching for IR excesses in 
Sun-like stars observed by {\em WISE}}
\author[F. Cruz-Saenz de Miera, M. Chavez, E. Bertone and O. Vega]{F. Cruz-Saenz de Miera\thanks{E-mail:
fsaenz@inaoep.mx (FCSM); mchavez@inaoep.mx (MC); ebertone@inaoep.mx (EB), ovega@inaoep.mx (OV)}, M. Chavez, 
E. Bertone and O. Vega\\
Instituto Nacional de Astrofisica, Optica y Electronica, Luis Enrique Erro 1, Tonantzintla, Puebla, Mexico}
\begin{document}

\date{Accepted . Received; in original form}

\pagerange{\pageref{firstpage}--\pageref{lastpage}} \pubyear{2013}

\maketitle

\label{firstpage}

\begin{abstract}
We present the results of a search of infrared excess candidates in a comprehensive 
(29\,000 stars) magnitude limited sample of dwarf stars, spanning the spectral range F2-K0, 
and brighter than V$=$15~mag. We searched the sample within the {\em WISE} all sky survey database 
for objects within 1 arcsecond of the coordinates provided by SIMBAD database and 
found over 9\,000 sources detected in all {\em WISE} bands. This latter sample excludes 
objects that are flagged as extended sources and those images which are affected by various optical artifacts.
For each detected object, we compared the observed W4/W2 (22$\mu$m/4.6$\mu$m) flux ratio with the expected 
photospheric value and identified 197 excess candidates at 3$\sigma$. For the vast majority of candidates, 
the results of this analysis represent the first reported evidence of an IR excess. Through the comparison 
with a simple black-body emission model, we derive estimates of the dust temperature, as well as of the dust 
fractional luminosities. For more than 80\% of the sample this temperature is higher than 120~K, 
suggesting the presence of warm circumstellar dust.
Complementary observations at longer wavelengths (far-IR and sub-mm) are required for better characterising 
and explaining the origin of this emission.
\end{abstract}


\section{Introduction}

An infrared excess (IR) around mature stars, measured as the flux in excess to that expected purely from the
stellar photosphere, is produced by orbiting dust particles. This dust is destroyed by physical processes such
as Poynting-Robertson drag, radiation pressure and photoevaporization in timescales much shorter than the stellar age
and therefore, it must be replenished by collisions of planetesimals.

Since the unexpected discovery of an IR excess around Vega \citep[$\alpha$ Lyrae,][]{aumann84}, 
infrared observations from space confirmed that the presence of circumstellar discs around main sequence 
(MS) stars is a common phenomenon \citep{trilling08}. However, these objects
show a large distribution in dust temperature ($T_{\rm dust}$) and mass and, while it is common to find cold 
debris discs ($T_{\rm dust} < 120$~K), warm discs around this type of stars are indeed unusual and have 
been detected for a few stars \citep[see, e.~g.,][]{kennedy12}. In spite of their rarity, there are strong 
motivations for their study and they have posed interesting theoretical challenges to explain their presence 
and relatively large masses. On the one hand, warm material is expected to be close to the host star, within 
a few astronomical units, therefore, they represent valuable laboratories to study the potential correlation 
between the detected material and the presence of larger bodies (planets) in orbits similar to that of the Earth. 
On the other hand, their fractional luminosities, well above the estimated fractional luminosity of the solar 
system zodiacal dust cloud \citep{dermott02}, escape explanations 
through the models of terrestrial planet formation \citep{kenyon05} or the steady state planetesimal grinding, 
and have to be formed through other mechanisms such as transient events \citep{wyatt07}.
Two main approaches have been suggested to explain the origin of hot dust: one of them considers a temporary 
increment in collisions due to dynamical instabilities, causing a so-called late heavy bombardment. Such a scenario 
has been proposed by \citet{lisse12} to account for the debris disc around $\eta$ Corvi. 
The second mechanism is the catastrophic collision of two rocky, 
planetary-scale bodies, which could explain the debris disc around BD+20~307 \citep{weinberger11}. Therefore, 
the study of warm debris discs not only is important to understand the dynamical evolution and
the presence of small bodies in planetary systems, but also
to understand the formation of terrestrial planets and the history of our solar system.

In this context, the recent Wide-field Infrared Survey Explorer ({\em WISE}) mission \citep{wright10} 
is certainly providing a fresher benchmark for characterising warm circumstellar material around main 
sequence stars. {\em WISE} has conducted an all sky survey with an unprecedented sensitivity (orders of magnitude 
better than its early predecessor {\em IRAS}) in four bands: 3.4, 4.6, 12, and 22 $\mu$m, denoted as W1, W2, W3, 
and W4, respectively. {\em WISE} data, together with the outcomes from {\em AKARI} 
\citep[see, e.g.,][]{fujiwara09,fujiwara13}, represent valuable means for demographic studies of warm 
circumstellar material. To date, most stellar analyses involving {\em WISE} data had 
focused on stellar samples aimed at searching for IR excesses in stars observed by the {\em Kepler} mission, 
including stars that are known to host transiting planetary companions \citep{krivov11,lawler12,ribas12,kennedy12}, 
and identifying possible correlations of mid-IR excesses and rotation \citep{mizusawa12}. 

In this paper we present a study of the {\em WISE} data base with the goal of searching excesses at 22~$\mu$m 
within a comprehensive magnitude-limited sample of Sun-like stars. The resulting set of excess 
candidates and the parameters derived for the circumstellar material will serve as a basis for future studies at 
longer wavelengths, which may shed light on the frequency and properties of warm circumstellar discs around this 
kind of stars.

\section{The stellar sample and {\em WISE} data}
The sample was constructed through a selection process that started by querying the SIMBAD database for main 
sequence stars. We imposed a limiting magnitude of V$=$15~mag, luminosity class V, and spectral types between 
F2 and K0. We also required 2MASS data to be available. From this first selection we got 
the information of nearly 29\,000 stars. It is important to note that the query by criteria in SIMBAD results in 
a wide variety of stellar sources with the vast majority (90\%) corresponding to objects whose object type is 
{\it only} an asterisk (*), this is, stars which do not present or has not been identified any peculiarity 
(binarity, variability, etc.). The other most prominent stellar groups within this sample are stars in binary 
or multiple systems (5\%), pre-main sequence objects (2\%), stars in clusters and associations (2\%), and the 
remaining corresponding to a wide variety of stellar sources, mainly variable stars of different types.

\begin{figure*}
\centering
\begin{minipage}{140mm}
\includegraphics[width=140mm]{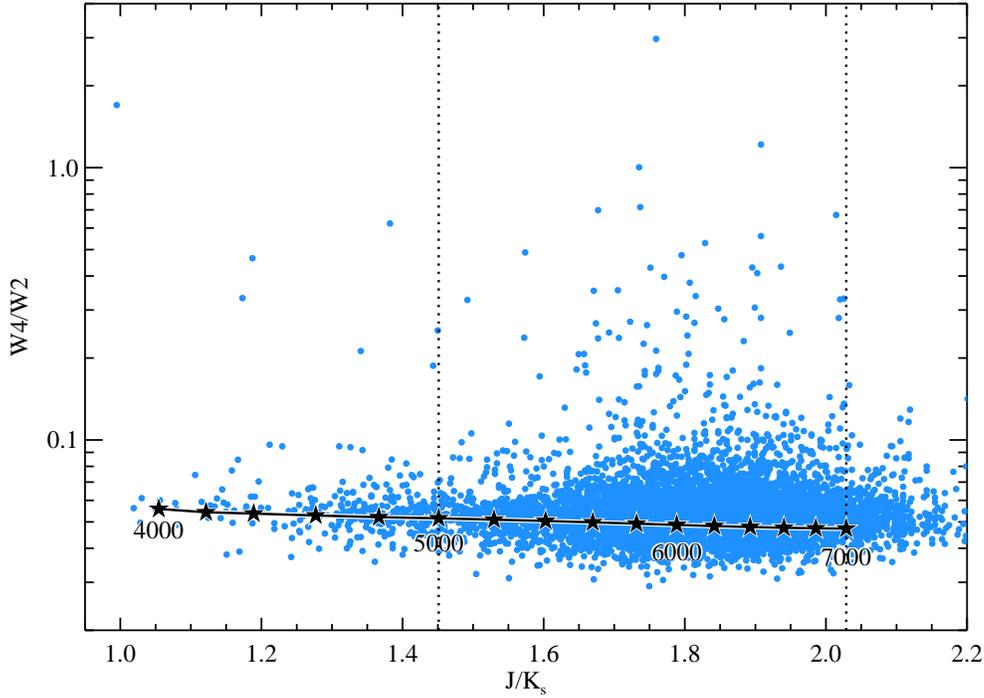}
\caption{Flux ratios diagram using {\em WISE} data at the W2
and W4 bands and the 2MASS fluxes at J and K$_{S}$. Blue dots represent the
detected stars classified as main sequence and brighter than V$=$15~mag in SIMBAD
database. The black stars indicate the position of the photospheric flux ratios calculated from theoretical 
fluxes extracted from the NEXTGEN library \citep{hauschildt99}. The vertical dashed lines indicate the approximate 
temperature boundaries for K0 (5\,000~K) and F2-type (7\,000~K) stars.}
\label{fig1}
\end{minipage}
\end{figure*}

{\em WISE} is a NASA funded explorer mission that was launched at the end of 2009 and conducted an all sky survey in 
four IR bands, with an angular resolution of 6.1, 6.4, 6.5, and 12.0~arcsec for W1, W2, W3, and W4, respectively. 
In spite of having slightly more than a third of the collecting area of its most immediate 
predecessor {\em AKARI}, its high performance detectors (with 4 million pixels) pushed down the limiting flux density 
to about  6~mJy at 22~$\mu$m, which is nearly 20 times better than that of {\em AKARI} at 18~$\mu$m \citep{ishihara10,wright10}.  
The {\em WISE} all sky data release\footnote{http://wise2.ipac.caltech.edu/docs/release/allsky/} contains information of 
more than 500 million objects, half of which are expected to be stellar sources, with an astrometric accuracy of better 
than 0.5~arcsec even for faint sources. The enhanced WISE sensitivity has provided 
and will provide in years to come access to a number of previously unidentified mid-IR excesses around solar-like stars.

We used the stellar coordinates to identify {\em WISE} counter parts. The constraints we imposed for the 
selection were {\em WISE} sources to be located within one arcsecond in radius and that reported fluxes at all 
{\em WISE} bands have signal-to-noise ratios (SNR) larger than 5. Additionally, 
following the criteria described in \citet{lawler12}, we have also excluded objects whose {\em WISE} data, in one or more 
bands, appear flagged as having contamination of halos, diffraction spikes, and optical ghosts due to nearby bright 
sources, as well as those flagged as extended sources. We have also excluded objects that are labelled as saturated 
in any of the W2, W3, and W4 bands, i.e. some of our objects after this selection might be saturated in the W1 band. 
After this process the sample was reduced to about 9\,500 detected objects.

Even though we do not expect interstellar extinction to be important for many objects in this later set, 
we have applied an extinction correction at the J, K$_{S}$ and W2 bands to those objects for which we identified reddening in excess 
of E(B-V)=0.05~mag. Reddening has been estimated from the comparison of the B-V color available in SIMBAD and the color 
index--spectral type calibration of \citet{pecaut12}. The extinction curve used for this procedure was that of
\citet{rieke85}. For the W4 band no extinction correction was applied, since at this wavelength the interstellar
reddening is expected to have negligible effects.

Detections are presented in Fig.~\ref{fig1}, where we plot the {\em WISE} flux ratio 
W4/W2 versus the 2MASS flux ratio J/K$_{S}$, for the entire sample. As a reference 
we have overplotted (connected stars) the theoretical photospheric line constructed from 
NEXTGEN model fluxes \citep{hauschildt99} in the effective temperature ($T_{\rm eff}$) range 
4\,000-7\,000~K with the step  available in the 
model flux library of 200~K. All theoretical fluxes are of solar chemical composition and surface 
gravity $\log{g}$=4.5~dex. Note that most data points clearly cluster around the photospheric line, and that point 
clustering extends well beyond these effective temperature limits, particularly at the lower temperature 
edge. This extension is most probably due to the low temperature ($T_{\rm eff}<5\,000$~K) members of 
binaries and multiple systems, both identified or probably still unidentified, and to reddened objects.
Note also that numerous objects in Fig.~\ref{fig1} are located well away from the photospheric line, 
hence plausibly corresponding to objects with prominent excesses.

\section{Finding IR excess candidates}
In order to determine if there is an excess and its quantification, we measured the difference 
between the observed flux ratios W4/W2 and theoretical photospheric predictions for a given 
J/K$_{S}$ and compared with the corresponding error $\sigma_{\rm tot}$. Errors in our data points have 
been calculated by considering three different sources: photometric errors $\sigma_{\rm obs}$, 
calibrations uncertainties $\sigma_{\rm cal}$,  and errors associated with the photospheric parameters 
$\sigma_{\rm phot}$. The $\sigma_{\rm obs}$ was obtained from the quoted uncertainties available within the {\em WISE}
data release for each band. Absolute calibration uncertainties of {\em WISE} are estimated to be 
2.4, 2.8, 4.5 and 5.7\% for the W1, W2, W3 and W4, respectively \citep{jarrett11}.
For $\sigma_{\rm phot}$ we considered the effects of varying surface gravity ($3.5 \leq \log{g} \leq 5.0$) 
and metallicity ($-3.5 \leq {\rm [M/H]} \leq 0.0$), once an effective temperature is assigned to each 
star by matching its J/K$_{S}$ color with the NEXTGEN theoretical fluxes.
The total error budget is then  $\sigma_{\rm tot} = \sqrt{\sigma_{\rm phot}^2 + \sigma_{\rm obs}^2 + \sigma_{\rm cal}^2}$. 
For the vast majority of the detected objects $\sigma_{\rm obs}^2$ is the dominant source of uncertainty.

We have considered an excess whenever $({\rm W4/W2}_{\rm WISE}-{\rm W4/W2}_{\rm phot})/\sigma_{\rm tot} \geq 3$, 
where ${\rm W4/W2}_{\rm WISE}$ is the extinction corrected observed color, and ${\rm W4/W2}_{\rm phot}$  is 
the theoretical photospheric flux ratio calculated using the
appropriate transmission functions and color corrections \citep{wright10}. 
Through this analysis we obtained a total of 313 excess candidates. At this stage it is necessary 
to elaborate in some cautionary notes on the reliability of these excesses.

\section{{\em WISE} images examination and final list of excess candidates}

\begin{figure}
\includegraphics[width=84mm]{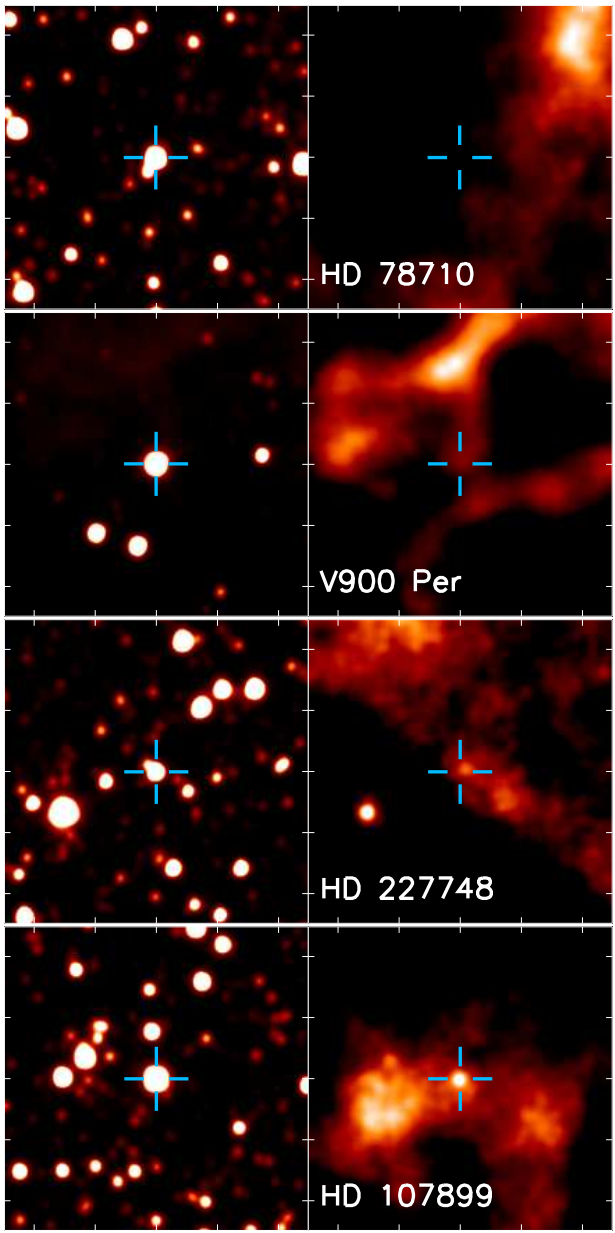}
\caption{WISE W2 (left panels) and W4 (right panels) 5$\times$5 arcmin images of HD~78710, V900~Per, 
HD~227748, and HD~107899; north is up and east is left. The color scale spans from 0 to 20 times the 
standard deviation of the background for W2, while for W4 the interval is reduced to 0--3$\sigma$. The three upper 
panels show objects that we classified as non-detections in W4, while HD~107899 is included in our catalog.
\label{fig2}}
\end{figure}

As a first step to assess on the quality of the excess candidates, we have carefully visually inspected 
images at the four bands for each of the 313 objects. Figure~\ref{fig2} summarises the results 
of this examination. We found numerous spurious detections in which no object is evident in the 
W4 band. An example of this situation is illustrated in the top two panels of the figure for the star
HD~78710. The object is clearly present in W2 (left) and absent in W4 (right). This effect is a product of the 
source extraction in the {\em WISE} pipeline processing and the complex backgrounds that characterise 
W4 images. In the second row of panels we show another case, exemplified by the star V900~Per, in which 
infrared emission at 22~$\mu$m is detected, but cannot be associated with a point source. In the third row we display the 
case of HD~227748 where there is an offset in the position of the source identified in W2 with respect to 
that in W4, and, in any case, the faint source cannot be unambiguously distinguished from emission peaks of the 
interstellar medium.

Objects in the above described cases (116 stars) were rejected leaving the final sample with 197 objects. 
In the panels at the bottom of Fig.~\ref{fig2} we illustrate an excess detection case for HD~107899. 
The star is clearly visible in the W4 band, even if it appears located within a region with difusse emission. 
However, many detections correspond to well isolated point sources.

\begin{figure}
\includegraphics[width=84mm]{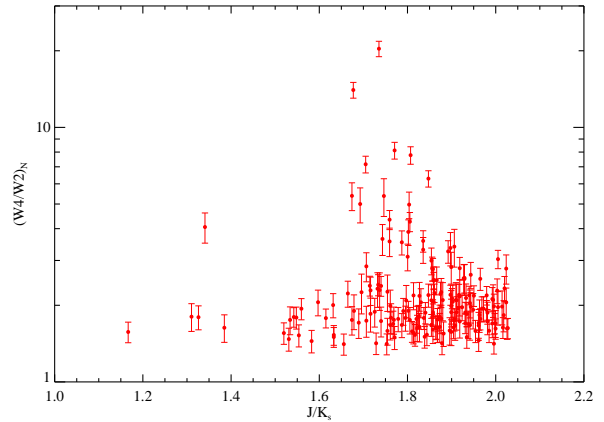}
\caption{Normalised flux ratios (W4/W2)$_{N}$ for the fiducial set of objects with excesses at 3$\sigma$.
\label{fig3}}
\end{figure}

In Fig.~\ref{fig3} we illustrate the distribution of sources with excesses at W4. The y-axis gives the
flux ratios (W4/W2)$_{N}$ normalised to the expected photospheric values versus J/K$_{S}$. 
The largest excess corresponds to the F5V type HD~39415, which is the only {\em IRAS} source in our set 
of excess candidates. This object was not detected at 60 and 100~$\mu$m and its flux at 25~$\mu$m was 
labeled as of moderate quality. Note that in the determination of excesses we have implicitly considered 
that the 2MASS fluxes and the flux at W2 are purely photospheric, as assumed in the analysis of \citet{krivov11}. 
Nevertheless, this assumption might not be correct for the few young stellar objects included in the sample. 
The object-type statistics for the 3$\sigma$ sample closely resembles that of the original sample from 
SIMBAD: isolated field stars account for the majority (about 90\%), while the remaining objects belong to various 
stellar types (stars in double systems, pre-main sequence stars, etc.). 

\begin{figure*}
\begin{center}$
\begin{array}{cc}
\includegraphics[width=84mm]{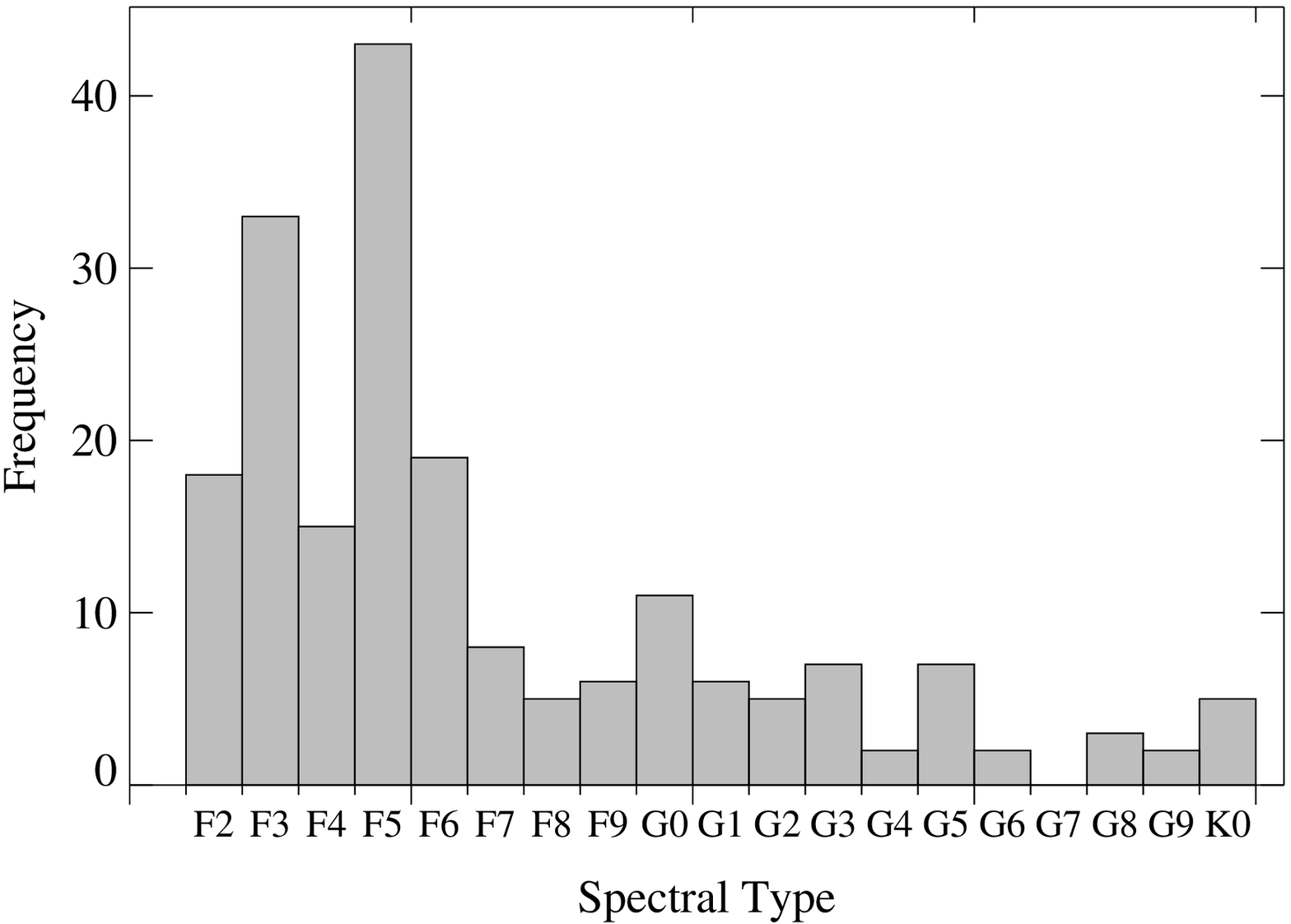} &
\includegraphics[width=84mm]{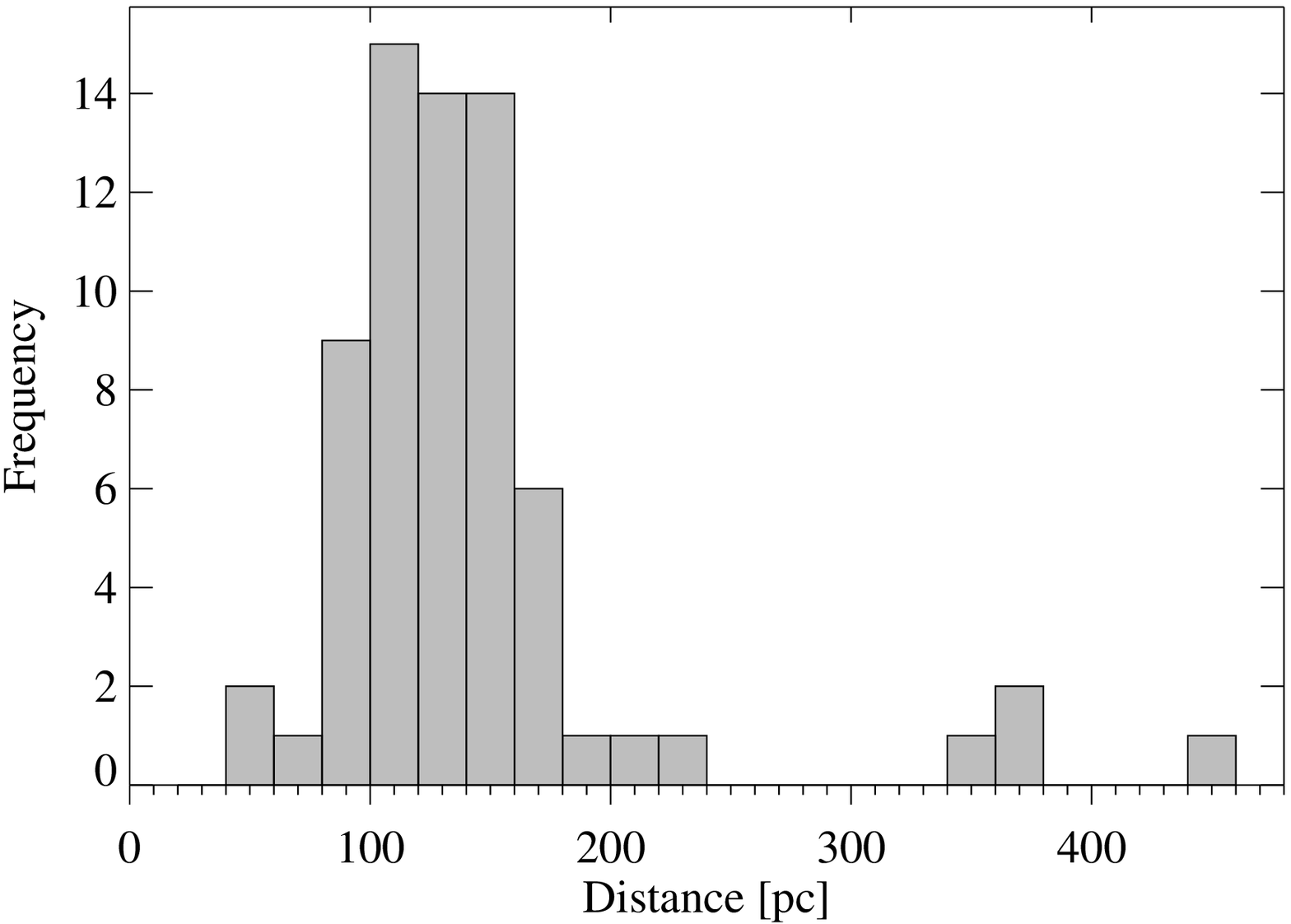}
\end{array}$
\end{center}
\caption{On the left we show the spectral type distribution of the 197 objects that display IR excesses. 
The distance distribution for the 68 objects with available Hipparcos parallaxes is shown on the right. 
In the distance histogram the bins are of 20~pc.}
\label{fig4}
\end{figure*}

In left panel of Fig.~\ref{fig4} we present the spectral type distribution of the excess candidates, which is
clearly dominated by the F-type stars, while in the right panel of that figure we show distance distribution for 
68 objects with available {\em Hipparcos'} data. In this group, most of the 3$\sigma$ excesses  are found in objects 
closer than 300~pc. There are, however, four objects farther away. Note 
the absence of the nearest stars (with distances less than  47~pc) in the histogram. The reason for this is 
explained by the adopted selection requirements for searching within the {\em WISE} database, as nearer objects 
are sources flagged as extended, saturated, and/or affected by one or more of the above mentioned 
optical artifacts in the W2 band.

\section{Discussion and Conclusions}
We have identified a sample of 197 main-sequence objects that display 3$\sigma$ IR excesses, many of which are present 
at both W3 and W4 {\em WISE} bands. This number corresponds to about 2\% of the parent sample of {\em WISE} detections.
In our resulting sample of IR excesses, 20 have previously reported observations in the mid-IR. The majority of these 
objects belong to the Scorpius-Centaurus association and have been studied with {\em Spitzer} \citep{chen11} and 
{\em WISE} \citep{rizzuto12}. There are also some Pleiades members, one of which is HD~23514, amongst the dustiest 
systems  in this cluster \citep{rhee08,sierchio10}. In the particular case of 
{\em WISE} data, \citet{krivov11} report an excess of one target, TrES-2, that fulfills our selection criteria. 
However, this object has not been included in our analysis because the SNR in the W4 band is below 
our adopted value of 5.

\begin{figure}
\includegraphics[width=84mm]{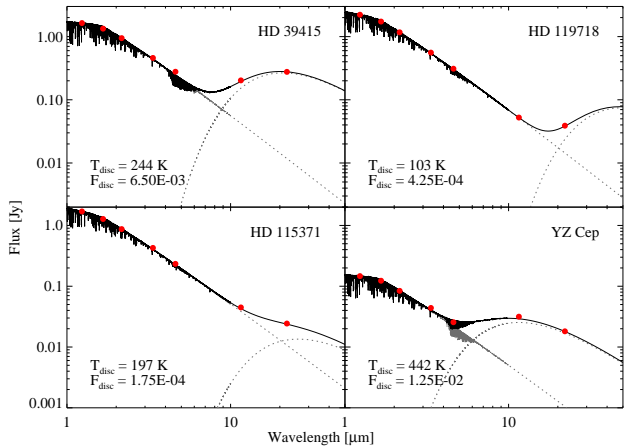}
\caption{Spectral energy distributions of four stars that display prominent 22$\mu$m excesses. 
Observational data points (red dots) mark the J, H, and K$_{S}$ 2MASS magnitudes, extracted from SIMBAD, and four  {\em WISE} bands. 
The continuous lines correspond to the best fit of the stellar photosphere plus a black body. In each panel we indicate the black body temperature and fractional luminosity that delivers the best fit.
\label{fig5}}
\end{figure}

In Fig.~\ref{fig5} we display the spectral energy distributions of four stars in our sample of excess candidates. 
Data points in red correspond to the fluxes of both 2MASS and {\em WISE} bands. The theoretical fluxes for each star have 
been calculated by interpolating within the NEXTGEN grid for the effective temperatures derived from their J/K$_{S}$ ratios. 
The solid line in each panel represents the star$+$disc spectral energy distribution (SED). This integrated SED was obtained 
following three steps. First, the theoretical fluxes were scaled in J to match the observational data points. Second, 
we created an extensive grid of black body flux densities for temperatures ($T_{\rm disc}$) in the range
20-500~K (at a step of 1~K), and fractional luminosities ($F_{\rm disc}$)
\citep{wyatt08},  in the interval 10$^{-1}$-10$^{-5}$. Third, we conducted
a $\chi^{2}$ minimization that provided the combination of $T_{\rm disc}$ and $F_{\rm disc}$ that best fit {\em WISE} bands. 
The best fit values are indicated in each panel of Fig.~\ref{fig5}.

In Table~\ref{tab1}, we list some of the sample stars for which we identified a prominent IR excess. 
The table provides the stellar identification, the full set of object types available in SIMBAD, the V magnitude 
and spectral types as given in SIMBAD, the stellar effective temperature and its 
uncertainty derived from the quoted errors in the J and K$_{S}$, the normalised flux ratios 
W3/W2$_{N}$ and W4/W2$_{N}$ associated with  the SNRs at the W3 and W4 bands. 
In the last two columns, we report the estimates for the dust temperature and fractional luminosities and their 
uncertainties, that were obtained fitting the {\em WISE} fluxes$\pm \sigma_{\rm tot}$. 
A label after the identification indicates those objects included in previous IR observational programs.
The table with the full stellar sample is available in electronic form.

\begin{figure}
\includegraphics[width=80mm]{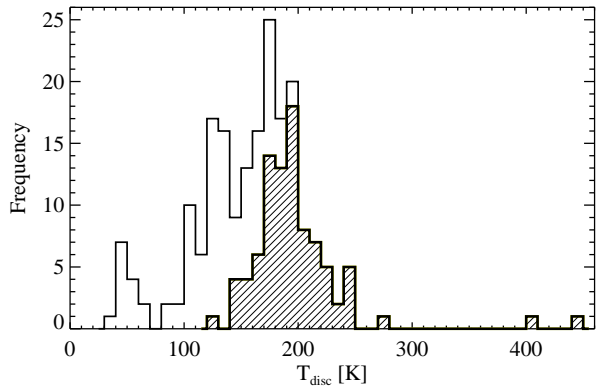}
\caption{Distribution of the black body temperature of the disc candidates of our sample. The shaded area indicates the
distribution of those objects that also display a 3$\sigma$ excess in W3.}
\label{fig6}
\end{figure}

\begin{figure}
\includegraphics[width=80mm]{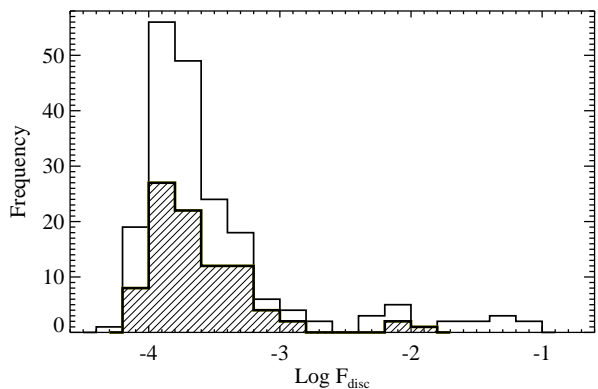}
\caption{Distribution of the fractional luminosity of the disc candidates of our sample. The shaded area indicates the
distribution of those objects that also display a 3$\sigma$ excess in W3.}
\label{fig7}
\end{figure}

In Fig.~\ref{fig6} we plot the distribution of the black body temperature associated to the 197 disc candidates of our sample. The majority of the stars (161 out of 197) have warm circumstellar material with $T_{\rm disc} > 120$~K, with a peak around 180~K; this result was expected because of the wavelength range covered by the {\em WISE} observations.  About 46\% (90) of the stars also show 3$\sigma$ excesses in W3 indicated with the shaded area in the figure. Note that this later subsample clearly occupies the hotter end of the distribution. In these cases we have at least two non-photospheric points available to determine the black body properties, hence the fitting procedure provides better estimates of the disc parameters. In general, the relative errors of both $T_{\rm disc}$ and  $F_{\rm disc}$ in objects with W3 and W4 excesses are lower than in objects displaying an excess in only W4. In contrast, the distribution of fractional luminosities (see Fig.~\ref{fig7}) does not indicate any particular 
trend for stars displaying also a W3 excess. The vast majority of excess stars present a disc fractional luminosity in the interval $10^{-3}-10^{-4}$. However, ten sources have a very strong IR excess, with $F_{\rm disc}> 10^{-2}$, and nine of them appear to have a very low temperature estimate ($T_{\rm disc} \leq 69$~K). The star YZ~Cep, also has a large fractional luminosity, and exhibits the hottest disc of our sample ($T_{\rm disc}=442$~K; see Fig.~\ref{fig5}). YZ~Cep is an irregular variable of RW Aurigae type \citep{ross26} and it is classified as a T~Tauri star \citep{kardopolov87}. Another object that presents a hot disc is HD~23514. Our modelling provides a dust temperature of 405~K which is significantly lower than 750~K obtained by \citet{rhee08}. This difference is most probably due to the fact that we conducted the fit at longer wavelengths (W3, W4), while  \citep{rhee08} fit the data up to $\lambda=12$~$\mu$m.

\begin{figure}
\includegraphics[width=84mm]{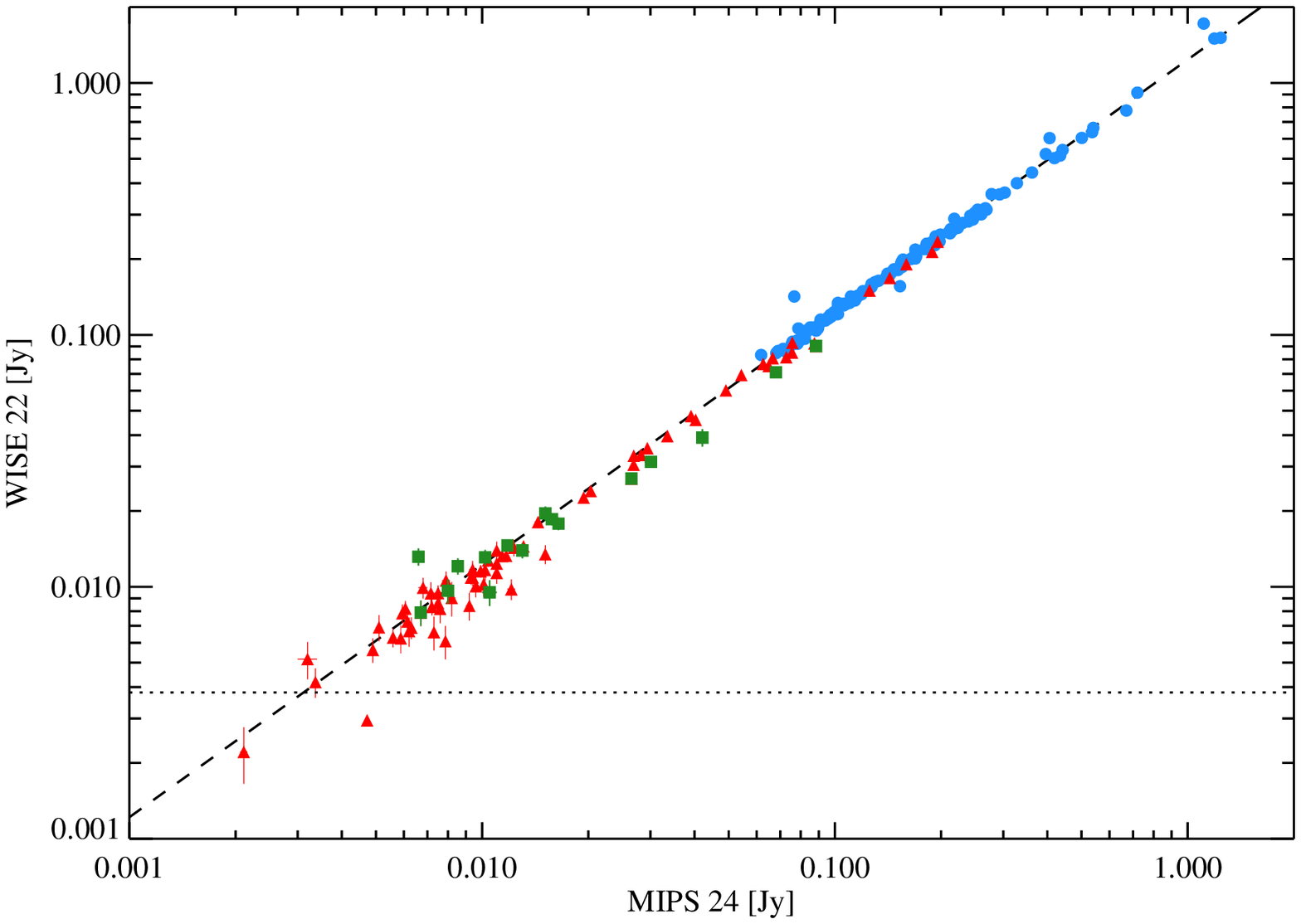}
\caption{Comparison between {\em WISE}-W4 and {\em Spitzer}-MIPS at 24~$\mu$m. Blue circles and red triangles are sources from
\citet{eiroa13} and \citet{chen11}, respectively. Data from \citet{chen11} consist of the 65 stars in their sample 
that accomplished  the {\em WISE} quality selection criteria adopted in this work. Green squares are the excess candidates 
included in the present work.
Error bars are smaller that the symbol size for most stars. The agreement of the two instruments is evident down to the lowest W4 flux level in our sample of excess candidates indicated with the dotted line. 
\label{fig8}}
\end{figure}

In the analysis here presented, we have examined {\em WISE} images to identify false detections. Such scrutiny turned out to be of fundamental importance since, as we have demonstrated, nominal high quality fluxes in the {\em WISE} database do not always correspond to a detected source. The lack of this double-check process yielded contrasting results in searches for IR excesses in stars of the {\it Kepler} field \citep{ribas12}. To further check on the reliability of our reported excesses we have compared {\em WISE}-W4 fluxes with those collected with {\em Spitzer}-MIPS at 24~$\mu$m. In  Fig.~\ref{fig8}  we plot this comparison for three different stellar sets. Squares in green correspond to the stars in our sample with previously reported IR data. Dots in blue are the fluxes reported in \citet{eiroa13} for the bright stars included in the Herschel-DUNES program \citep{eiroa10}. The red triangles indicate the position of stars in \citet{chen11}. For this latter set we have extracted {\em WISE} data 
considering the same criteria as for our working sample. It is interesting to see that there is a remarkable consistency down to the weakest source in our sample (3.8~mJy for TYC 9139-2239-1), whose flux level is indicated with the horizontal dotted line.

A mid-IR excess indicates, in general, the existence of warm circumstellar material. Apart from the few very hot objects, at these wavelengths we are mapping 
the Wien-edge of the energy distributions, therefore, the lack of an excess does not necessarily imply the absence of 
circumstellar material. Cool discs ($T_{\rm dust}<50$~K) have their imprints at longer wavelengths, and very little or no excess in 
the mid-IR. The results of this work incorporate numerous newly identified objects with mid-IR excesses and
will certainly motivate new observational programs at far-IR, sub-mm and mm that are required to better characterise 
the material around Sun-like stars.

\begin{table*}
  \caption{Subsample of stars with prominent mid-IR excesses. \label{tab1}}
  \begin{tabular}{llrcrrrrrll}
  \hline
 star ID  & O-types  &  V  & Sp. Type & $T_{\rm eff}$ (K) & $\frac{W3}{W2}_{N}$ & SNR$_{\rm W3}$ &  $\frac{W4}{W2}_{N}$  & SNR$_{\rm W4}$ & $T_{\rm disc}$ (K) & $F_{\rm disc}$ \\
 \hline
HD 15053  & *,IR &  9.50 & F7V & 6610$\pm$250 &  1.29 &  2.11 &  1.67 &  3.95 &  166$^{+42}_{-30}$ &  9.25E-05$^{+3.25E-05}_{-1.75E-05}$ \\
HD 23380  & *,IR &  8.38 & F2V & 6870$\pm$330 &  1.28 &  3.10 &  1.61 &  8.19 &  183$^{+26}_{-24}$ &  7.50E-05$^{+7.50E-06}_{-5.00E-06}$ \\
HD 39415$^{\dagger}$  & *,IR &  8.40 & F5V & 5810$\pm$370 &  5.18 & 50.05 & 20.37 & 45.18 &  244$^{+1}_{-4}$   &  6.50E-03$^{+0.00E+00}_{-2.50E-04}$ \\
HD 65516  & *,IR &  8.91 & F5V & 6210$\pm$210 &  1.31 &  3.68 &  1.53 &  7.00 &  205$^{+36}_{-27}$ &  1.00E-04$^{+0.00E+00}_{-0.00E+00}$ \\
HD 115371 & *,V*,IR &  8.29 & F3V & 6600$\pm$290 &  1.41 &  7.35 &  2.19 & 13.22 &  197$^{+16}_{-15}$ &  1.75E-04$^{+0.00E+00}_{-0.00E+00}$ \\
HD 119718$^{\dagger}$ & *,IR &  8.03 & F5V & 6440$\pm$250 &  1.23 &  0.26 &  2.65 &  7.75 &  103$^{+44}_{-59}$ &  4.25E-04$^{+6.46E-02}_{-2.00E-04}$ \\
HD 125344 & *,IR &  9.78 & F7V & 6180$\pm$240 &  1.25 &  0.88 &  3.58 & 10.64 &  105$^{+18}_{-52}$ &  7.25E-04$^{+1.93E-02}_{-2.25E-04}$ \\
HD 141227 & *,IR &  8.85 & F5V & 5710$\pm$170 &  1.73 & 14.63 &  7.16 & 24.76 &  164$^{+6}_{-2}$ &  1.50E-03$^{+0.00E+00}_{-0.00E+00}$ \\
HD 158815 & *,IR &  9.70 & G1V & 5900$\pm$300 &  1.51 &  7.08 &  4.33 & 14.21 &  163$^{+9}_{-10}$ &  7.50E-04$^{+7.50E-05}_{-5.00E-05}$ \\
HD 196813 & *,IR &  9.72 & G3V & 5990$\pm$280 &  1.41 &  4.30 &  3.53 &  8.48 &  147$^{+15}_{-12}$ &  5.75E-04$^{+1.00E-04}_{-7.50E-05}$ \\
HD 204765 & *,pr*,IR &  9.91 & G1V & 6040$\pm$210 &  1.31 &  2.15 &  3.10 &  6.88 &  133$^{+24}_{-16}$ &  4.50E-04$^{+1.50E-04}_{-1.00E-04}$ \\
YZ Cep    & *,V*,RI*,IR & 11.40 & G0V & 5820$\pm$180 &  8.77 & 62.90 & 14.54 & 17.98 &  442$^{+58}_{-5}$  &  1.25E-02$^{+0.00E+00}_{-2.50E-03}$ \\
\hline
\end{tabular}

\begin{flushleft}$^\dagger$ Star with previous mid-IR observations\end{flushleft}
\end{table*}

\section*{Acknowledgments}

We thank the anonymous referee for constructive comments that improved the presentation
of this work. MC and FCSM would like to thank CONACyT for financial support through grant number 134985. This publication makes use of data products from the Wide-field Infrared Survey Explorer, which is a joint project of the University of California, Los Angeles, and the Jet Propulsion Laboratory/California Institute of Technology, funded by the National Aeronautics and Space Administration. 
This research has made use of the SIMBAD database, operated at CDS, Strasbourg, France.


\bsp

\label{lastpage}

\end{document}